\documentclass[10pt,aps,pra,twocolumn,floatfix,superscriptaddress]{revtex4-1}
\usepackage{amsmath,amssymb,amsfonts,bbm,graphicx,color,mathtools}

\usepackage[breaklinks, colorlinks=true, urlcolor=blue, anchorcolor=blue,%
 citecolor=blue, filecolor=blue, linkcolor=blue, menucolor=blue, %
 linktocpage=true, pdfproducer=medialab, pdfa=true]{hyperref}

\usepackage[utf8]{inputenc}
\usepackage[T1]{fontenc}
\usepackage{physics}
\usepackage{pdfcomment}
\usepackage{newtxtext,newtxmath}

\newcommand{\suc}{\text{succ}}
\begin{document}
\title{Quantum spatial search on graphs subject to dynamical noise}
\author{Marco Cattaneo}
\affiliation{Quantum Technology Lab,
Dipartimento di Fisica ``Aldo Pontremoli'', Universit\`a degli Studi di
Milano, I-20133 Milano, Italy}
\author{Matteo A. C. Rossi}
\affiliation{QTF Centre of Excellence, Turku Centre for Quantum Physics,
Department of Physics and Astronomy, University of Turku, FI-20014 Turun Yliopisto, Finland}
\author{Matteo G. A. Paris}
\affiliation{Quantum Technology Lab,
Dipartimento di Fisica ``Aldo Pontremoli'', Universit\`a degli Studi di
Milano, I-20133 Milano, Italy}
\author{Sabrina Maniscalco}
\affiliation{QTF Centre of Excellence, Turku Centre for Quantum Physics,
Department of Physics and Astronomy, University of Turku, FI-20014 Turun Yliopisto, Finland}
\affiliation{QTF Centre of Excellence, Department of Applied Physics,
School of Science, Aalto University, FI-00076 Aalto, Finland}
\date{\today}
\begin{abstract}
We address quantum spatial search on graphs and its implementation by continuous-time quantum walks in
the presence of dynamical noise. In particular, we focus on search on the complete graph and on the star graph of
order N, also proving that noiseless spatial search shows optimal quantum speedup in the latter,
in the computational limit $N\gg 1$. The noise is modeled by independent sources
of random telegraph noise (RTN), dynamically perturbing
the links of the graph. We observe two different behaviors depending on the
switching rate of RTN: fast noise only slightly degrades performance,
whereas slow noise is more detrimental and, in general, lowers the
success probability. In particular, we still find a quadratic speed-up
for the average running time of the algorithm, while for the star graph
with external target node we observe a transition to
classical scaling. We also address how the effects of noise depend on
the order of the graphs, and discuss the role of the graph topology.
Overall, our results suggest that realizations of
quantum spatial search are possible with current
technology, and also indicate the star graph as the perfect candidate
for the implementation by noisy quantum walks, owing to its
simple topology and nearly optimal
performance also for just few nodes.
\end{abstract}
\maketitle
\section{Introduction}
Quantum spatial search \cite{Aaronson2003} is the problem of finding a marked element in a structured database, i.e. a database whose items are connected by a structure of links mimicking a graph.
Essentially, it is the generalization of the Grover algorithm \cite{Grover1996} to search problems in which one has to take into account the spatial organization of the dataset.

Childs and Goldstone showed that an algorithm based on continuous-time quantum walks (CTQWs) \cite{Farhi1998trees} may solve the problem of quantum spatial search on certain graph topologies in a time $T={O}(\sqrt{N})$ \cite{Childs2004}, where $N$ is the order of the graph, thus outperforming any classical algorithm, where
the searching time is bounded to $T={O}(N)$.
In particular, they proved that the full speed-up of order $T={O}(\sqrt{N})$ is achieved in the case of the complete graph, the hypercube graph and the $d$-dimensional lattice for $d\geq 4$.

In recent years, many other graph topologies have been considered.
For instance, the algorithm has been investigated on complete bipartite graphs \cite{Novo2015}, on balanced trees \cite{Philipp2016a}, on Erd\"{o}s-Rényi graphs \cite{Chakraborty2016,Glos2018}, on the simplex of the complete graph \cite{Wong2016b} and on graphs with fractal dimensions \cite{Agliari2010,Li2017}.
Moreover, it has been shown that high connectivity and global symmetry of the graph are not necessary for fast quantum search \cite{Janmark2014,Meyer2015}. The first result of this paper is a proof of the optimality of quantum spatial search on the star graph, both when the target is the central node and when it is one of the external ones.
\par
These results are very promising, but in order to address concrete
implementations, one should consider the presence of noise and disorder
in the system. In particular, one should analyze the effect of noise
on the success probability of the algorithm and on the scaling of the
searching time. As a matter of fact, the study of the effects of noise
on spatial search is still at the early stages. The robustness against
 noise upon considering adiabatic quantum computation has been studied
\cite{Roland2005}, as well as the performance of spatial search on graphs
with broken links \cite{Novo2015}. More recently, it has been shown that
the coupling to a thermal bath may improve the efficiency of the algorithm
in the presence of static disorder \cite{Novo2017}, whereas a fully-dynamical description of the noise is still missing. The search algorithm has been
analysed on random temporal networks \cite{Chakraborty2017}, i.e.
Erd\"{o}s-Rényi graphs whose topology changes after a certain time
interval, though this model can hardly mimic the dynamics of real noise.
\par
In this paper, we address continuous-time quantum spatial search on
graphs subject to dynamical noise. In particular, we analyze the
performance of the algorithm on the complete graph and on the star
graph, after having analytically proven that the search is optimal also
on the latter. The noise is modeled as random telegraph noise (RTN)
affecting the links of the graph with tunable strength, ranging from a weak
perturbation of the hopping amplitudes to a strength comparable to the coupling, inducing dynamical percolation.
Our choice for the noise is motivated by its relevance in
systems of interest for quantum information processing \cite{Galperin2006,Abel2008,Joynt2011,Rossi2016,Benedetti2013a},
and by the fact that RTN is at the root of the $1/f$ noise affecting
superconducting qubits \cite{Paladino2014a}. In recent years some works
have addressed the properties of CTQWs on the one-dimensional lattice
subject to random telegraph noise \cite{Benedetti2016,Siloi2017,Siloi2017a,Piccinini2017}, also in
the presence of spatial correlations \cite{Rossi2017a}.
In this paper we analyze the effects of RTN on spatial search on
graphs with generic topology. Other models of CTQW subject to dynamical
noise have been proposed as well \cite{Yin2008,Amir2009,Darazs2013}.
\par
The paper is structured as follows: in Sec.~\ref{sec:algorithm} we review the continuous-time quantum spatial search algorithm and we prove its optimality on the star graph.
In Sec.~\ref{sec:noiseModel} we introduce the noise model and we discuss the noisy evolution of the walker.
In Sec.~\ref{sec:completeGraph} we present our results on the effects of noise on the complete graph, while in Sec.~\ref{sec:starGraph} we focus on the star graph.
Sec.~\ref{sec:concludingRemarks} closes the paper with some concluding remarks.
\section{The algorithm}
\label{sec:algorithm}
Given a certain graph $G$ composed of $N$ nodes, we want to find the marked element $w$, called target node.
The graph $G$ is described by the adjacency matrix $A$, whose elements are defined as
\begin{equation}
A_{ij} = \begin{cases}
  1 & \text{if nodes $i,j$ connected} \\
  0 & \text{otherwise}.
\end{cases}
\end{equation}
The Hilbert space of the walker is $\mathcal{H}=\text{span}\{\ket{j}\}$ with $j=1,\ldots,N$, where $\ket{j}$ is the single-particle localized state associated to the node $j$.
The Hamiltonian of the algorithm reads
\begin{equation}
\label{eqn:Hamiltonian}
H=\gamma L + H_w = \gamma L-\ket{w}\!\!\bra{w},
\end{equation}
where $H_w=-\ket{w}\!\!\bra{w}$ is called \textit{oracle Hamiltonian}, $\gamma$ is a suitable coupling constant and we introduced the Laplacian matrix
$L = D - A$, where $D$ is the degree matrix, a diagonal matrix where the $i$-th entry is the degree of the $i$-th node, i.e. the number of links connected to it. Notice that we are neglecting an overall constant in $H$, which fixes the unit of measure for time $t$ and related quantities.

The quantum walk starts in the fully delocalized state $\ket{s}$, where
\begin{equation}
\ket{s}=\frac{1}{\sqrt{N}}\sum_{j=1}^N \ket{j},
\end{equation}
and the state at time $t$ reads
\begin{equation}
\ket{\psi(t)}=e^{-iHt}\ket{s}.
\end{equation}
After the time $t$, we measure in the vertices basis. The probability of measuring the walker in the target node is
$p_w(t) = \abs{\braket{w}{\psi(t)}}^2$.
We define the success probability $p_\suc$ as the maximum probability
\begin{equation}
  p_\suc = \max_t p_w(t),
\end{equation}
and $T$ the smallest time instant for which $p_\suc$ is achieved.
Optimizing the search algorithm then consists in finding $\gamma$
such that for $T$ as small as possible the success probability
$p_\suc=$ is maximal. We say that the algorithm is optimal
if $p_\suc\approx 1$ in a time $T=O(\sqrt{N})$.
\par
As an example, we review the performance of the algorithm on the complete graph, which had already been addressed employing a different computational framework as the ``analog analogue'' of Grover's algorithm \cite{Farhi1998}.
The action of the Hamiltonian on the states $\ket{s}$ and $\ket{w}$, using as $L$ in Eq.~\eqref{eqn:Hamiltonian} the Laplacian of the complete graph and choosing $\gamma=1/N$, reads
\begin{equation}
H\ket{s}=-\frac{1}{\sqrt{N}}\ket{w},\qquad H\ket{w}=-\frac{1}{\sqrt{N}}\ket{s}.
\end{equation}
Therefore, the Hamiltonian drives transitions between the two states and after a time $T=\pi\sqrt{N}/2$ we have $\ket{\psi(T)}=\ket{w}$, i.e. the algorithm is optimal for any order $N$.
\subsection{Optimality of the star graph}
Let us now address the proof of the optimality of the algorithm on the star graph.
The search on the star graph has already been investigated in \cite{Novo2015} as a particular case of complete bipartite graphs, and a success probability $p_\suc\approx 1/2$ in a time $T=O(\sqrt{N})$ was found.
However, in \cite{Novo2015} the authors choose to use the adjacency matrix instead of the Laplacian operator in the Hamiltonian of the algorithm Eq.~\eqref{eqn:Hamiltonian}.
This choice is irrelevant if the graph is regular, as the elements on the diagonal of $L$ are all equal and are thus just a energy shift, but it leads to completely different dynamics in non-regular graphs \cite{Wong2016}, as is the case with the star graph.

In what follows, we prove that the continuous-time quantum spatial search is optimal on the star graph, if we employ the Laplacian as in Eq.~\eqref{eqn:Hamiltonian}.
The star graph consists of $N-1$ nodes connected to a central node.
There are two different situations to consider: the case in which the target node $w$ is the
central one, and the case in which the target node is one of the external nodes.

Let us start with the case in which the target node is the central node of the star graph, named $\ket{c}$.
If $\ket{c}=\ket{w}$, by choosing $\gamma=1/N$ we obtain
\begin{equation}
H\ket{s}=-\frac{1}{\sqrt{N}}\ket{w},\qquad H\ket{w}=-\frac{1}{\sqrt{N}}\ket{s},
\end{equation}
therefore the dynamics is analogous to the one on the complete graph and we find the target node with $p_\suc=1$ after $T=\pi\sqrt{N}/2$, independently of the graph order $N$.
\par
The proof of the optimality when the target is one of the external nodes is
more involved, and it is extensively addressed in the Appendix.
By making use of the Krylov subspace method to reduce the space of
the walker \cite{Novo2015}, and then, by employing degenerate perturbation theory \cite{Janmark2014}, we show that, in
the computational limit $N\gg 1$, $\ket{\psi(T)}=\ket{w}+O(N^{-1/2})$ at time $T=\pi\sqrt{N}/2$, i.e. the algorithm is optimal.
Notice that in this case we have to choose $\gamma = 1$ for the algorithm to succeed.
\begin{figure}[h!]
  \includegraphics{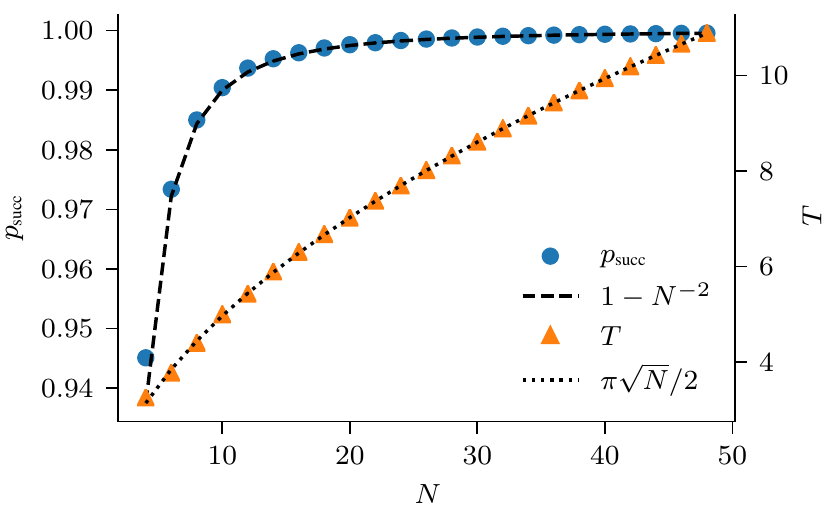}
  \caption{(Color online) Success probability $p_\suc$ (blue circles, left axis) and optimal time $T$ (orange triangles, right axis) of quantum spatial search on the star graph as a function of the order $N$ with an external node as target. The dots represent the exact  quantities, the dotted line is the benchmark $ \pi \sqrt{N} / 2$ and the dashed line represents the function that better approximates $p_\suc$, which we have found to be $1-N^{-2}$, showing that the remainders in Eq.~\eqref{eqn:evolutionStateStar} are actually of order $N^{-1}$, rather than $N^{-1/2}$.
The plot shows that, even for finite $N$, the success probability is very high and that the optimal time follows the asymptotic behavior.}
  \label{fig:maxProbTimeStarGraph}
\end{figure}
\par
While the proof shows the optimality of the algorithm for large $N$, Fig.~\ref{fig:maxProbTimeStarGraph} shows that the success probability is close to $1$ also for small values of the order $N$, $p_\suc \approx 1 - N^{-2}$,  and the optimal time scales as $\sqrt{N}$.
This suggests that the star graph may be a good candidate for an experimental implementation of continuous-time quantum spatial search, since just few
nodes are required to achieve quantum speed-up. Furthermore, the star graph
has a simpler topology compared to the other graphs that are suitable for
spatial search, thus it might be easily realized in a laboratory using e.g. superconducting circuits.
\section{The noise model}
\label{sec:noiseModel}
The random telegraph noise (RTN) is the continuous-time stochastic
process that describes the dynamics of a bistable fluctuator, i.e. a
quantity which switches randomly between two given values (say $\pm 1$)
according to a certain switching rate $\mu$.
The RTN is completely characterized as $\{g(t),\, t \in[0,+\infty)\}$
where $g(t) = \pm 1$, which implies that the probability of
switching $n$ times in a time $t$ follows a
Poisson distribution
\begin{equation}
p_\mu(n,t)=e^{-\mu t}\frac{(\mu t)^n}{n!}.
\end{equation}
The stochastic process is stationary and its autocorrelation
function reads
\begin{equation}
\langle g(\tau)g(0)\rangle=e^{-2\mu\abs{\tau}},
\end{equation}
corresponding to a Lorentzian spectrum.

Motivated by the kind of noise
observed in superconducting networks,
we model the environmental noise by assuming that the links of the graph
are affected by independent and equal (i.e. with the same switching
rate $\mu$) RTN. Accordingly,  we modify the Laplacian operator in Eq.~\eqref{eqn:Hamiltonian}, keeping the classical probability conservation
rule for which the sum of the elements in a column of the Laplacian matrix
is zero.

The noise is described by the $N\times N$ matrix $\mathbf{g}(t)$,
where $N$ is the number of nodes in the graph and $g_{jk}(t)$ is the
stochastic process describing the noise on the link connecting $j$ to $k$.
The matrix $\mathbf{g}(t)$ is thus symmetric, zero-diagonal and has
only $l$ independent entries, where $l$ is the number of links in the graph. Since the noises on different links
are independent of each other, we have, for the non-zero entries of $\mathbf{g}(t)$,
\begin{equation}
\langle g_{jk}(\tau)g_{j'k'}(0)\rangle=e^{-2\mu\abs{\tau}}
(\delta_{jj'}\delta_{kk'}+\delta_{jk'}\delta_{kj'})\,.
\end{equation}

We now replace Eq.~\eqref{eqn:Hamiltonian} with a
noisy Hamiltonian depending on the stochastic process $\mathbf{g}(t)$.
The noisy Laplacian $L^{(\mathbf{g})}(t)$ in the node basis reads
\begin{equation}
\label{eqn:noisyHam}
L^{(\mathbf{g})}_{jk}(t)=\begin{cases}
-\left[1+\nu g_{jk}(t)\right]&\textnormal{ if }(j,k)\textrm{ connected}\\
D_{jk}+\nu \sum_{i=1}^Ng_{ik}(t)&\textnormal{ if }j=k \\
0&\textnormal{ otherwise,}
\end{cases}
\end{equation}
where $\nu$ is the noise strength.
If $\nu=1$ we obtain dynamical percolation, i.e. the random creation
and removal of links in the graph according to the switching rate $\mu$.
The Hamiltonian then reads
\begin{equation}
  H^{(\mathbf{g})}(t) = \gamma L^{(\mathbf{g})}(t) - \ket{w}\!\!\bra{w}\,.
\end{equation}

If the initial state of the walker is $\rho_0=\ket{s}\!\!\bra{s}$,
the evolved density matrix is the ensemble average
\begin{equation}
\label{eqn:evolution}
\rho(t)=\langle U(t)\rho_0 U(t)^\dagger\rangle_{\{\mathbf{g}(t)\}},
\end{equation}
where $\langle \ldots\rangle_{\{\mathbf{g}(t)\}}$ denotes the average over all possible realizations of the stochastic process $\mathbf{g}(t)$ and $U(t)$ is the unitary evolution operator associated to a particular realization, given by
\begin{equation}
U(t)=\mathcal{T}\exp{-i\int_0^t ds\,H^{(\mathbf{g})}(s)},
\end{equation}
where $\mathcal{T}$ is the time-ordering operator.

Eq.~\eqref{eqn:evolution} defines a map that describes the dynamics of the open quantum system, and $\rho(t)$ is the only relevant physical quantity for investigating the evolution of the system. From this point of view, the noise model discussed above is just an effective microscopic description of the coupling between system and environment that generates the quantum map that we are actually observing.

The success probability at time $T$ is now the matrix element
\begin{equation}
p_\suc=\mel{w}{\rho(T)}{w}.
\end{equation}
\begin{figure}[t]
\includegraphics{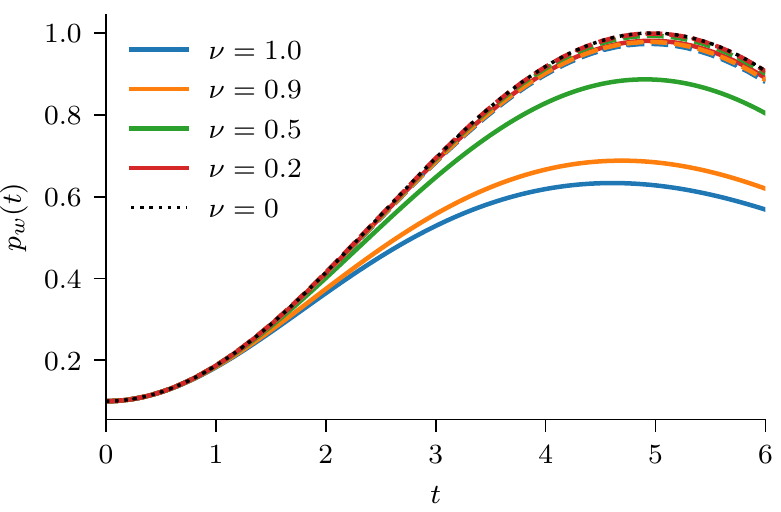}%
\caption{(Color online) Probability of measuring the target node $p_w(t)=\mel{w}{\rho(t)}{w}$ as a function of time $t$ on the complete graph of order $N=10$, for fast noise ($\mu=10$, dashed lines) and slow ($\mu=0.01$, solid lines).
Different values of the noise strength $\nu=0.2,0.5,0.9$ and $1.0$ are shown respectively, from top to bottom, by the red, the green, the orange and the blue lines. The dotted black line describe the noiseless case ($\nu = 0$). Fast noise barely affects the algorithm, while slow noise decreases the success probability, but also the optimal time.}
\label{fig:probFtimeN10Complete}
\end{figure}

\begin{figure*}
  \includegraphics{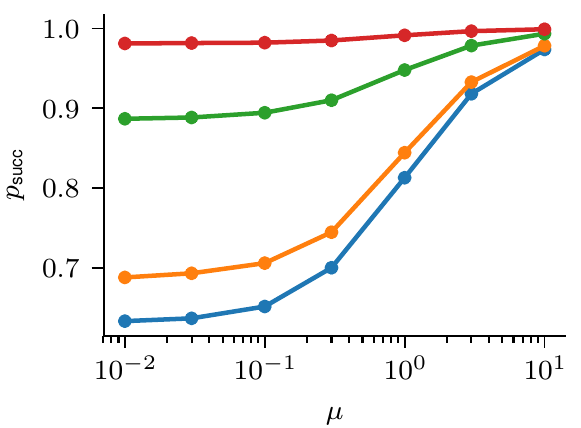}%
  \includegraphics{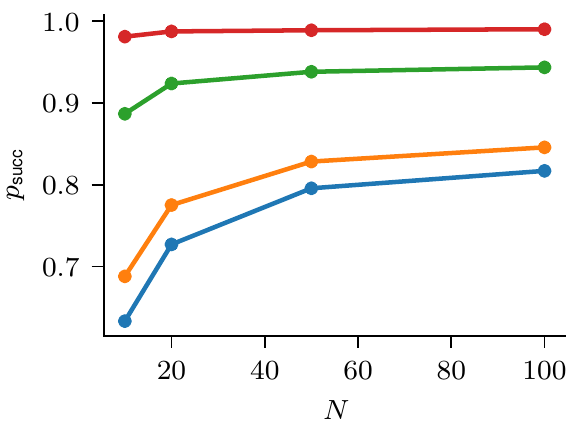}%
  \includegraphics{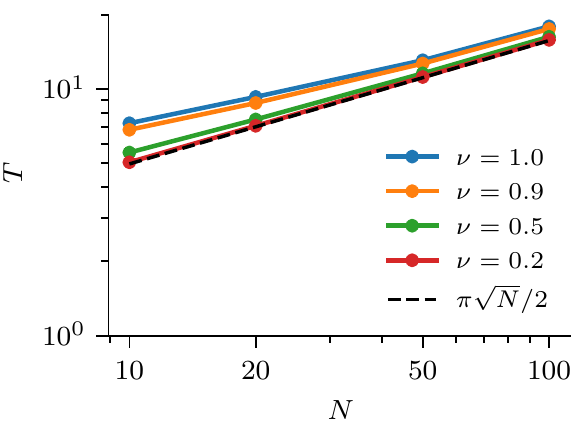}
\caption{(Color online) The left panel shows the success probability $p_\suc$ of the quantum spatial search on the complete graph of order $N=10$, as a function of the switching rate $\mu$. The central panel and right panel show, respectively, the success probability and the average running time $T$ on the complete graph as a function of the order $N$ for slow noise ($\mu=0.01$). In the three plots, different values of the noise strength $\nu=0.2,0.5,0.9$ and $1$ are shown respectively (from top to bottom in the first two panels, from bottom to top in the last one), by the red, green, orange and blue lines. The success probability decreases with the noise strength, but it increases with $N$. Although the success probability is lower than $1$, the average running time is proportional to $\sqrt{N}$ (showed with a black dashed line), thus the algorithm can still outperform a classical one for sufficiently large $N$.}
\label{fig:probSuccFNComplete}
\end{figure*}

\section{Quantum spatial search on noisy graphs}
\label{sec:results}
In this section we discuss how random telegraph noise affects
continuous-time quantum spatial search on the complete and the star graphs.
\subsection{Complete graph}
\label{sec:completeGraph}
The time evolution of the walk, given by Eq.~\eqref{eqn:evolution}, cannot be computed analytically for a large number of links, therefore we simulate the dynamics numerically and then we average over a big number of realizations of the noise.
Since this is the optimal value in the noiseless case, we set $\gamma=1/N$ in Eq.~\eqref{eqn:noisyHam}.
The code used in this work is written in Julia \cite{Bezanson2017} and is available on GitHub \cite{QuantumSpatialSearch}. Since the number of noise trajectories explored in the simulation is finite, fluctuations are present on the mean value leading to the quantum map Eq~\eqref{eqn:evolution}. We have calculated the standard deviation of this mean value and considered a number of noise realizations that is large enough to make such standard deviation irrelevant, i.e. non-visible in the graphs. In particular, after this careful analysis of numerical uncertainties, we have chosen to average over $10 000$ realizations of the noise when $\mu \geq 1$, and over $20\,000$ when $\mu < 1$.

In order to analyze the robustness of the search in the presence of noise, we focus on the success probability of the algorithm and on the optimal time $T$.
We explore several scenarios by varying three fundamental parameters: the order of the graph $N$, the noise strength $\nu$, and the switching rate $\mu$ of the RTN.
In particular, we identify two different regions of values of the switching rate, and we call the RTN with $\mu \lesssim 1$ \textit{slow} or \textit{semi-static noise}, and the RTN with $\mu \gtrsim 1$ \textit{fast noise}.
In Fig.~\ref{fig:probFtimeN10Complete} we plot the probability of measuring the target node $p_w(t)=\mel{w}{\rho(t)}{w}$ as a function of time, for $N=10$ and choosing $\mu=10$ and $\mu=0.01$.
Several values of the noise strength $\nu$ are considered.

\begin{figure*}
\includegraphics{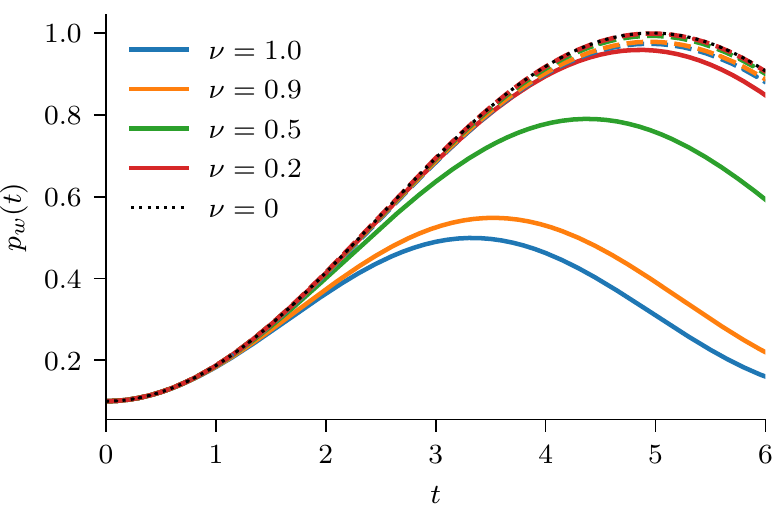}
\includegraphics{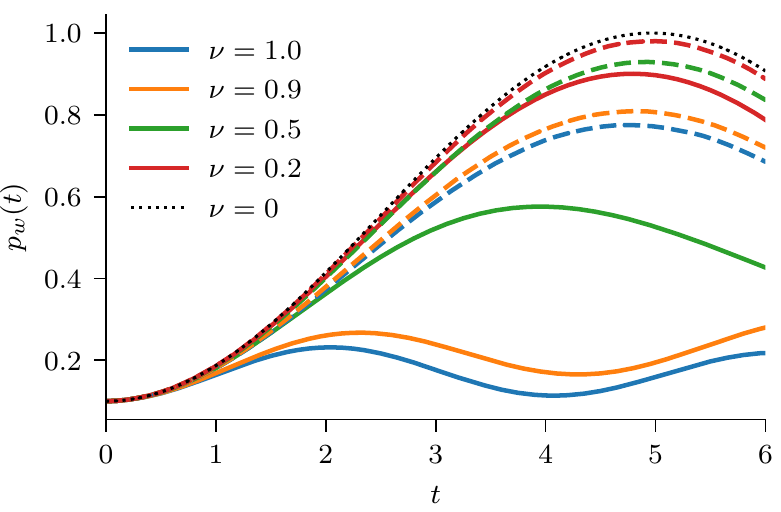}
\caption{(Color online) Probability of measuring the target node $p_w(t)=\mel{w}{\rho(t)}{w}$ as a function of time, on the star graph of order $N=10$ with central node  (left) and an external node (right) as target, for slow noise (solid lines, $\mu=0.01$) and fast noise (dashed lines $\mu=10$).
Different values of the noise strength $\nu=0.2,0.5,0.9$ and $1$ are represented by the red, green, the orange and the blue lines (from top to bottom), respectively.
The dotted black lines describe the noiseless case. If the target node is the central one, the results are qualitatively similar to the case of the complete graph, but the effect of noise is stronger. In the case of an external target node, slow noise dramatically affects the success probability.}
\label{fig:probFtimeN10Star}
\end{figure*}

A clear difference in the behaviours appears, depending on the value of the switching rate $\mu$: for fast noise the algorithm is still optimal, in the sense that we obtain a success probability (i.e. the maximal probability of measuring the target node) $p_\suc\approx 1$ in a time $T\approx \pi\sqrt{N}/2$; on the contrary, slow noise significantly affects the efficiency of the search, and for $\mu=0.01$ and $\nu=1$ the probability of success is around $60\%$.
At any time, the probability of measuring the target node for a fixed switching rate and a certain noise strength $\nu$ is always lower than for a smaller noise strength, proving that in general the presence of dynamical noise jeopardizes the algorithm.
The left panel of Fig.~\ref{fig:probSuccFNComplete} depicts the success probability as a function of the switching rate $\mu$, showing that decreasing the switching rate of the noise leads to worse and worse performance.

For higher orders of the graph we have obtained qualitatively similar results, although increasing the order leads to slightly better success probabilities.
This is intuitive, since by adding nodes to the complete graph we are creating more possible paths connecting each node to the target, decreasing the effects of broken links due to semi-static noise.
The success probability as a function of the order $N$ is depicted in the central panel of Fig.~\ref{fig:probSuccFNComplete}, for slow noise and several values of $\nu$.
Further analysis suggests that changing the value of the coupling constant $\gamma$ in the presence of noise does not improve the results of the spatial search, but the optimal value remains $\gamma=1/N$ as in the noiseless case.

It should be noticed that, while the success probability tragically decreases in the presence of semi-static noise, the time $t_{\text{max}}$ at which we find the maximal success probability is slightly smaller.
From a computational point of view, one may assume the possibility of ``recognising'' the outcome of the spatial search algorithm, being able to tell whether or not it is the right solution.
In this framework we are allowed to run more trials of the algorithm, until the
correct solution is found. The probability of getting the right target node at the $n$-th trial is given by
\begin{equation}
\label{eqn:geometricDistribution}
p_g(n)=(1-p_\suc)^{n-1}p_\suc.
\end{equation}
Eq.~\eqref{eqn:geometricDistribution} is a geometric distribution with mean value $\langle n \rangle=1/p_\suc$.
Therefore, the average optimal time $T$ of the algorithm with success probability $p_\suc$ is given by
\begin{equation}
\label{eqn:avSuccessTime}
T=\frac{t_{\text{max}}}{p_\suc}.
\end{equation}

This is the time we should compare with the optimal analogue in the noiseless case. The right panel of Fig.~\ref{fig:probSuccFNComplete} shows the results for the average optimal time $T$ as a function of $N$, for slow noise and for several values of $\nu$.
Apart from a constant factor in the logarithmic scale, it is clear that the average optimal time still follows the quadratic speed-up $T\sim O(\sqrt{N})$, for any value of noise strength. With fast noise (not shown), the curves are closer to the noiseless case.

We stress, however, that this should not lead to underestimate the effect
of noise, since a success probability close to 1 is an important feature
in quantum spatial search. Indeed, besides being the analogue of Grover's
algorithm on structured databases, quantum spatial search may have other
applications/interpretations. For instance, let us consider a system
composed of quantum nodes connected by links, and let us assume to know that one of the nodes is affected by a certain potential well (mimicking the oracle
Hamiltonian), but without knowing where it actually is. In this case, we
may find the marked node by running the quantum spatial search
algorithm, but we would not be able to recognise the solution, unless
the success probability is close to unit.
\subsection{Star graph}
\label{sec:starGraph}
In this subsection we present the results about noisy quantum spatial search on the star graph.
We consider first the case in which the target is the central node, and then the case in which the target is one of the external nodes.
Indeed, the dynamics of the quantum walk in the two scenarios is remarkably different (for instance, in the former case we set $\gamma=1/N$ while in the latter $\gamma=1$), and, as we will see, the effect of the noise is different as well.
\subsubsection{Central target node}
In this case the effect of the dynamical noise on the search algorithm is similar to the case of the complete graph.
In the left panel of Fig.~\ref{fig:probFtimeN10Star} we plot the probability of measuring the target node as a function of time, for several values of $\nu$ and for both fast and slow noise.
The optimal $\gamma$ in the presence of noise still remains $\gamma=1/N$.
Qualitatively, we obtain the same results of the previous section: fast noise lightly influences the performance of the search, while slow noise is highly detrimental, although the success probabilities are slightly lower than in the case of the complete graph.

This is easily explained, for instance, in the semi-static scenario and percolation regime: at time $t=0$ around $50\%$ of the links of the star graph will be broken, and they will remain broken on average for almost all the evolution, since the noise is slow.
Each node, apart from the central one, has only one link, therefore is highly probable that the walker will remain ``stuck'' in the isolated nodes, and it will not find the target node.
On the contrary, in the case of the complete graph it is very unlikely that a certain node starts with all the links cut, therefore the walker will almost always find a path in the graph to reach the target.

\begin{figure}[t]
  \includegraphics{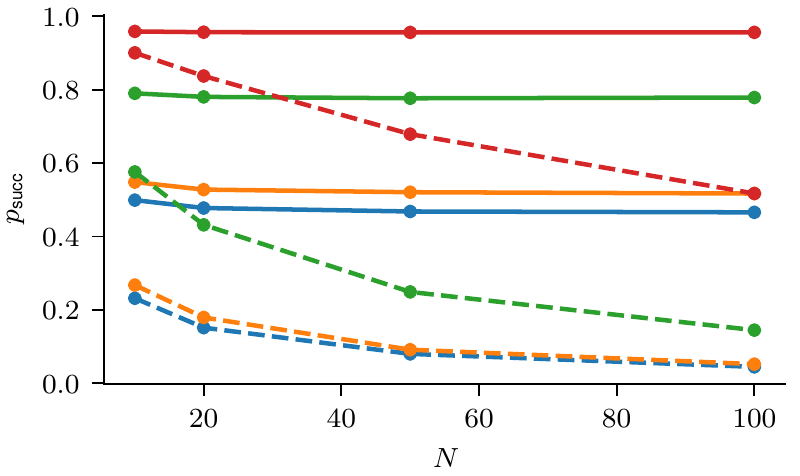}
  \caption{(Color online) Success probability on the star graph with central (solid lines) and external (dashed lines) target node, as a function of the order $N$, for slow noise ($\mu=0.01$).
  Different values of the noise strength $\nu=0.2,0.5,0.9$ and $1$ are shown, respectively, from top to bottom, by the red, green, orange and blue
   lines. While $p_\suc$ remains constant with $N$ if the target node is the central one, it vanishes when the target node is external, showing how the connectivity of the target node is relevant in the presence of noise.}
  \label{fig:psucc_vs_N_g001star}
\end{figure}

This phenomenon is independent of the order $N$, and this is the reason why, on the star graph, increasing the order does not lead to better results for semi-static noise, as depicted in Fig.~\ref{fig:psucc_vs_N_g001star} (solid lines).
However, the asymptotic behaviour of the average optimal time in the framework of iterated trials, given by Eq.~\eqref{eqn:avSuccessTime}, does not change, as shown in Fig.~\ref{fig:optTimeStar} (solid lines) only for slow noise; therefore, all the considerations discussed for the complete graph still hold.
\par
\subsubsection{External target node}
Here we analyze the effects of noise on the search on the star graph when
the target node is external. Numerical analyses have suggested that decreasing the value of $\gamma$ in the presence of slow noise may lead to better performance, while when we increase the switching rate the optimal $\gamma$ shifts toward the noiseless value $1$.
However, in this work we are interested in the effects of noise on the ideal algorithm, therefore we keep $\gamma=1$ in all the following analyses.

The results of the noisy algorithm on the star graph with external target node are shown in the right panel of Fig.~\ref{fig:probFtimeN10Star}.
Qualitatively we observe the same behaviour as for the central target node, although the success probability is quantitatively way worse than in the previous cases and even fast noise affects the performance in a non-negligible way.
This behaviour is not surprising, since the degree of the target node is only $1$, therefore if the noise, especially the semi-static one, affects the connecting link, then the target cannot ``exchange'' probability anymore with the rest of the graph.

Furthermore, the maximal probability of success decreases when the order of the graph increases.
This might be due to the fact that, as $N$ increases, the degree of the target node remains the same, while the degree of the central node, which is the only connection of the target to the rest of the graph, goes up as well, therefore the probability current may ``take the wrong direction'' more easily.

\begin{figure}[t]
\centering
\includegraphics{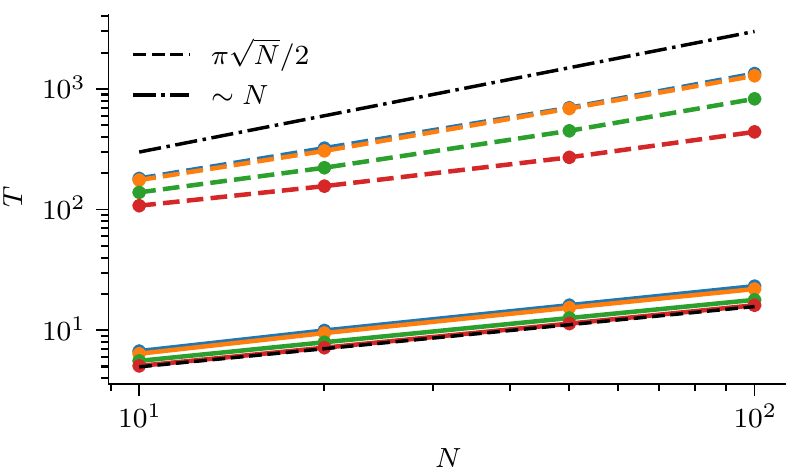}
\caption{(Color online) Average optimal time $T$ on the star graph with central (solid lines) and external (dashed lines) target node, as a function of the order $N$, for slow noise with $\mu=0.01$.
Different values of the noise strength $\nu=0.2,0.5,0.9$ and $1$ are shown, respectively, from bottom to top, by the red, green, orange and blue lines.
The dashed black describes the noiseless case $T= \pi\sqrt{N}/2$, while the dot-dashed black line shows a $T\sim N$ dependence. While in the case of the central target node the algorithm still shows a quantum speedup, if the target node is external the noise makes the algorithm transition to a classical scaling with $N$.}
\label{fig:optTimeStar}
\end{figure}

Finally, we investigate how the optimal time $T$ varies in the presence of noise.
Fig.~\ref{fig:optTimeStar} shows that in the case of slow noise ($\mu=0.01$) the scaling of the optimal time follows a transition from the quantum speed-up $T=O(\sqrt{N})$ to the classical time $T=O(N)$.
The case of the star graph with external target node is the only one in which the noise affects both the success probability and the optimal time, once again proving that this topology is particularly weak with respect to the effects of dynamical noise.
\section{Concluding remarks}
\label{sec:concludingRemarks}
Quantum spatial search via continuous-time quantum walks has received
much attention in the recent past. Several kinds of graphs have been
proposed, and their properties studied trying to understand if and how the
topology of the graph is correlated with fast spatial search.
In this paper we have taken a step further and we have proven the
optimality of the algorithm on the star graph, showing that the success
probability of the search is close to unit in a optimal time
$T\approx \pi\sqrt{N}/2$ also when the order $N$ of the graph is
low, regardless of the position of the target node.
This is particularly interesting since, owing to the simple topology
of the star graph and to the feasibility of the search employing just
few nodes, our results pave the way to the experimental
implementation of the continuous-time quantum spatial search algorithm.
\par
We have also addressed the performance of quantum spatial search in the
presence of dynamical noise. In particular, we have modeled the noise
as a collection of bistable fluctuators with the same switching rate
$\mu$, which induce independent random telegraph noise on each link
of the graph. We have studied the effects of noise in several scenarios,
e.g. by varying the order of the graph $N$, the switching rate $\mu$ and
the noise strength $\nu$, and we have analyzed it on the complete graph,
on the star graph with central node as target, and on the star graph with
one of the external nodes as target. Our results show that, in general,
the noise is detrimental for the probability of success of the search,
while it does not affect the quadratic speed-up of the time of the search $T=O(\sqrt{N})$, up to factors independent of $N$. This fact, however,
should not lead to underestimate the detrimental effect of noise,
since the success probability is the crucial quantity in any
search algorithm.
\par
Upon analyzing several noise scenarios, we have shown that the random
telegraph noise with large switching rate, i.e. fast noise, affects only
slightly the performance of the spatial search; in particular, it
decreases the success probability in a non-trivial way only when
applied on the star graph with one of the external nodes as target.
On the contrary, slow noise strongly jeopardizes the efficiency of
the algorithm.
\par
Finally, we have discussed how the topology of the graph plays a
role in the robustness against the dynamical noise, in particular
looking at the degree of the target node and at the connectivity
of the graph. The complete graph, having the maximal possible
connectivity and the maximal possible degree of the target node, is particularly resistant to the noise, and by increasing its order we obtain better results, since we are also increasing both the connectivity and the target degree.
The star graph with central node as target is slightly more affected by the slow noise, but increasing the order does not lead to better performance, since the connectivity of the graph remains the same.
The star graph with one of the external nodes as target has the lowest possible connectivity and target degree, and indeed the spatial search on it is heavily deteriorated by the presence of dynamical noise.
Increasing the order does not improve the algorithm, on the contrary it provides worse performance, since the target degree remains the same while the possible connections with the central node increase, opening more ``wrong ways'' for the probability current going toward the target node.
While connectivity seems to be irrelevant for noiseless quantum walks \cite{Janmark2014,Meyer2015}, our work points out that higher connectivity of the target node plays an important role in the presence of noise.
\par
Our analysis represents a step toward the understanding of the effects
of noise in continuous-time quantum spatial search. In particular, the
study of classical dynamical noise is important in view of implementing
the algorithm on a physical system which is unavoidably disturbed by
the external environment, as for the case of the superconducting qubits
subject to RTN and $1/f$ noise.
\begin{acknowledgments}
MC was supported by the EU through the Erasmus+ programme. MACR and SM
acknowledge support from the Academy of Finland via the Centre of
Excellence program (project 312058) and project 287750.
MGAP has been supported by JSPS through FY2017 program
(grant S17118) and by SERB through the VAJRA program
(grant VJR/2017/000011).
\end{acknowledgments}

\appendix*
\section{Proof of the asymptotic optimality on the star graph with external target node}
\label{app:proof}
Here we give the detailed proof that spatial search is optimal on the star graph, when one of the external nodes is the target.

Because of the high symmetry of the star graph, it is straightforward to see that the quantum walk is confined in the Krylov subspace given by the span of the vectors $\{\ket{c},\ket{w},\ket{s_{N-1}}\}$, where
\begin{equation}
\ket{s_{N-1}}=\frac{1}{\sqrt{N-2}}\sum_{\substack{j=1 \\ j\neq c,w}}^N \ket{j}.
\end{equation}
The reduced Hamiltonian in the above basis, choosing $\gamma=1$, reads
\begin{equation}
H_{\text{red}}=\begin{pmatrix}
N-1&-1&-\sqrt{N-2}\\
-1&0&0\\
-\sqrt{N-2}&0&1
\end{pmatrix}.
\end{equation}

We now extract a factor $N$ from the Hamiltonian, in order to employ degenerate perturbation theory; we will insert it again only at the end of the proof, when we will find the perturbed eigenvalues and eigenvectors.
We divide the Hamiltonian into two parts, $H^{(0)}$ and $H^{(1)}$, defined as
\begin{equation}
H^{(0)}=\begin{pmatrix}
1&0&-\sqrt{N-2}/N\\
0&0&0\\
-\sqrt{N-2}/N&0&0
\end{pmatrix},
\end{equation}
\begin{equation}
H^{(1)}=\begin{pmatrix}
-1/N&-1/N&0\\
-1/N&0&0\\
0&0&1/N
\end{pmatrix}.
\end{equation}
The overall Hamiltonian is given by $H_{\text{red}}=H^{(0)}+H^{(1)}$, up to the factor $N$.

We must be careful in employing perturbation theory, since we have to deal with two different orders, namely $O(N^{-1/2})$ and $O(N^{-1})$, and the second one is the square of the first one, thus the off-diagonal elements of $H^{(0)}$ cannot be neglected in a trivial way in the series expansion of the perturbation.
Therefore, we try to get to a better form of the Hamiltonian by diagonalizing $H^{(0)}$.

The eigenvalues of the Hamiltonian $H^{(0)}$ are
\begin{equation}
E^{(0)}_0=0,\qquad E^{(0)}_{1,2}=\frac{1\mp\sqrt{1+4/N-8/N^2}}{2},
\end{equation}
with associated eigenvectors respectively
\begin{align}
\ket{e_0}&=\ket{w} \notag \\
\ket{e_1}&=\mathcal{N}_1\left(-\sqrt{N}E^{(0)}_1\ket{c}+\ket{s_{N-1}}\right) \\
\ket{e_2}&=\mathcal{N}_2\left(-\sqrt{N}E^{(0)}_2\ket{c}+\ket{s_{N-1}}\right), \notag
\end{align}
where $\mathcal{N}_1$ and $\mathcal{N}_2$ are suitable normalization constants.
Notice that for $N\gg 1$
$\ket{e_1}=\ket{s}+O(N^{-1/2})$, and $E^{(0)}_1\sim N^{-1}$, therefore $E^{(0)}_0$ and $E^{(0)}_1$ are degenerate eigenvalues in the computational limit.
This is crucially important, since otherwise the dynamics would remain confined near to $\ket{e_1}$ at any time $t$, up to factors of order $O(N^{-1/2})$.
We have chosen $\gamma=1$ exactly to get $\ket{w}$ and $\ket{e_1}$ asymptotically degenerate.

Being careful about the orders of the perturbation, we can now use degenerate perturbation theory \cite{Sakurai}.
First of all, we rewrite $H_{\text{red}}$ in the new basis $\{\ket{w},\ket{e_1},\ket{e_2}\}$.

In the asymptotic limit $N\rightarrow\infty$, we obtain
\begin{equation}
H_{\text{red}}\sim \begin{pmatrix}
0&-N^{-3/2}&N^{-1}\\
-N^{-3/2}&N^{-2}&2N^{-3/2}\\
N^{-1}&2N^{-3/2}&1
\end{pmatrix}.
\end{equation}

We now diagonalize (up to factors of order $O(N^{-2})$) the $2\times 2$ matrix representing the subspace of the asymptotically degenerate eigenvectors $\ket{w}$ and $\ket{e_1}$.
Therefore, once again we change basis and we choose $\{(\ket{w}+\ket{e_1})/\sqrt{2},(\ket{w}-\ket{e_1})/\sqrt{2},\ket{e_2}\}$.

In this new basis, the total Hamiltonian reads
\begin{equation}
H_{\text{red}}\sim \begin{pmatrix}
-N^{-3/2}&-N^{-2}/2&N^{-1}/\sqrt{2}\\
-N^{-2}/2&N^{-3/2}&N^{-1}/\sqrt{2}\\
N^{-1}/\sqrt{2}&N^{-1}/\sqrt{2}&1
\end{pmatrix}.
\end{equation}

Eventually, we can use perturbation theory.
Indeed, the off-diagonal elements can at maximum bring a contribution of order $O(N^{-2})$ to the perturbed eigenvalues, while the diagonal elements are of order $O(N^{-3/2})$.
We still have off-diagonal elements in the submatrix of the asymptotically degenerate eigenvectors, but once again the contribution is of order $O(N^{-2})$.
Overall, this means that the ground state $\ket{\lambda_0}$ and the first
excited state $\ket{\lambda_1}$ of the Hamiltonian are
\begin{align}
  \ket{\lambda_0} &= (\ket{w}+\ket{e_1})/\sqrt{2}+O(N^{-1/2}) \\
  \ket{\lambda_1} &= (\ket{w}-\ket{e_1})/\sqrt{2}+O(N^{-1/2})
\end{align}

The contribution of order
$O(N^{-1/2})$ is brought by the off-diagonal elements in the submatrix of the asymptotically degenerate eigenvectors. The corresponding
eigenvalues (inserting again the factor $N$ we extracted at the beginning of the proof) are given by $E_0=-1/\sqrt{N}+O(N^{-1})$, $E_1=1/\sqrt{N}+O(N^{-1})$.

Therefore, in the computational limit $N\gg 1$ the evolution of the initial
state reads:
\begin{equation}
\label{eqn:evolutionStateStar}
\begin{split}
\ket{\psi(t)}&=e^{-iHt}\ket{s}=e^{-iHt}(\ket{e_1}+O(N^{-1/2}))\\
&=\frac{e^{-iHt}}{\sqrt{2}}\left(\ket{\lambda_0}-\ket{\lambda_1}+O(N^{-1/2})\right)\\
&=\frac{e^{-i E_0 t}}{\sqrt{2}}\left(\ket{\lambda_0}-e^{-i(E_1-E_0)t}\ket{\lambda_1}+O(N^{-1/2})\right)\\
&=\frac{e^{-i E_0 t}}{\sqrt{2}}\left(\ket{\lambda_0}-e^{-\frac{2it}{\sqrt{N}}+O(N^{-1})}\ket{\lambda_1}+O(N^{-1/2})\right).
\end{split}
\end{equation}

At the time $T = \pi \sqrt{N} / 2 $ we have $\ket{\psi(T)} = \ket{w} + O(N^{-1/2})$  i.e. the probability of success is one and the algorithm is optimal, as $T \propto \sqrt{N}$.

\bibliography{library}

\end{document}